\documentclass[twocolumn,prl,showpacs,preprintnumbers,amsmath,amssymb]{revtex4}

\usepackage{graphicx}
\usepackage{dcolumn}
\usepackage{bm}

\begin{document}
\title{\textbf{\textrm{Low prevalence, quasi-stationarity and power-law distribution in a model of spreading}}}

\author{Afshin Montakhab}
    \email{montakhab@shirazu.ac.ir}
\author{Pouya Manshour}

\affiliation{Department of Physics, College of Sciences, Shiraz University, Shiraz 71454, Iran}

\date{\today}
\begin{abstract}
Understanding how contagions (information, infections, etc) are spread on complex networks is important both from practical as well as theoretical point of view. Considerable work has been done in this regard in the past decade or so. However, most models are limited in their scope and as a result only capture general features of spreading phenomena. Here, we propose and study a model of spreading which takes into account the strength or quality of contagions as well as the local (probabilistic) dynamics occurring at various nodes. Transmission occurs only after the quality-based fitness of the contagion has been evaluated by the local agent. The model exhibits quality-dependent exponential time scales at early times leading to a slowly evolving quasi-stationary state. Low prevalence is seen for a wide range of contagion quality for arbitrary large networks. We also investigate the activity of nodes and find a power-law distribution with a robust exponent independent of network topology. Our results are consistent with recent empirical observations.
\end{abstract}


\maketitle
\section{Introduction}
Spreading, defined broadly, as transmission of contagions (e.g. information or virus) from one agent to the next is a fundamental process with application in a wide range of disciplines including physics, epidemiology and social sciences. In many cases of spreading phenomena, transmission only occurs after some local conditions are met. From a theoretical point of view, this requires construction of models of contagion dynamics (spreading) which take into account the content of the contagion as well as the role of the individual agents receiving/evaluating/transmitting contagion. It is therefore surprising that such models have not been proposed and studied, despite the fact that much attention has been given to spreading phenomena, typically in a mean-field approximation, in the past decade \cite{2,3,4,5}. In this Article, we propose and study a model of contagion spreading which takes into account the \textquotedblleft quality\textquotedblright of contagion as well as the role of the individual local agents in the spreading phenomenon. We find many interesting and realistic features in our model, including quality-dependent fast initial spreading followed by extremely slow dynamics, as well as robust power-law behavior in activities of individual agents. Low prevalence is also observed as a generic but limiting behavior of our model, e.g. as system size diverges.

It is hardly possible to overemphasize the importance of contagion dynamics. Modern telecommunications including emails and text-messages, and more recently social networks like \emph{Facebook} and \emph{Twitter} have revolutionized the process of information spreading. Network and/or viral marketing \cite{6}, opinion formation and rumor/innovation spreading \cite{7,8}, recruiting and talent searches \cite{9}, disaster response and relief efforts, mobilizing masses \cite{10}, as well as epidemic spreading \cite{2,3,11} are a few examples of how contagion spreading through complex networks is of vital importance in our modern way of life. Scarcity of reliable empirical results, as well as the ease to capture the general features of spreading, have led many authors to study spreading along the lines of epidemiology \cite{2,12,13,14}. In such approaches, one typically uses SI, SIR, or SIS models on a complex network where the letters in the acronyms refer to the state (Susceptible, Infected, or Recovered) of the agents on the network. Assuming a mean-field transmission rate $\lambda$, one typically finds a low threshold for spreading where a large part of the network is \textquotedblleft infected\textquotedblright in a relatively short time, with some dependence on network topology \cite{3,14,15,16}. However, such results are in contrast with most real-world observations which indicate a low prevalence in a variety of spreading phenomena \cite{17}. Here, we introduce the concept of contagion \textquotedblleft quality\textquotedblright in such models and take into account the role of individual local agents in evaluating transmission condition based on the observed \textquotedblleft fitness\textquotedblright of such contagion. We employ local probabilistic conditions for  transmission on two types of most commonly studied complex networks. We find that our model can provide a more realistic picture of spreading phenomena, leading to results consistent with general empirical observations of such phenomena.

\section{A model with contagion quality and local dynamics}
In most commonly studied models of spreading, the role of the active (local) agents is often reduced to a uniform (mean-field) behavior where upon receiving, the local agent transmits the contagion with probability $\lambda$ independent of the particular agent. Some important (but less-noticed) exceptions may be found here \cite{181,182,183}. Another shortcoming of such general models is the fact that the \textquotedblleft quality\textquotedblright or \textquotedblleft strength\textquotedblright of the contagion is often neglected. However, not all contagions are equal. For example, a new virus or a shocking news have a much more probability of transmission than a common virus or a normal everyday news. In fact, both these shortcomings of the most frequently studied models are related in a sense that in most spreading phenomena, local agents \textquotedblleft judge\textquotedblright the quality of the transmission received before deciding to pass it on. Typically, the contagion interacts with the local agent, and if some \textquotedblleft fitness\textquotedblright conditions are met, transmission occurs. To the best of our knowledge, models where local agents use the quality of the contagion they receive in order to decide whether or not to pass it on have not been studied before. In the present work, we introduce a quality factor for the contagion being spread and thus introduce local dynamics whereby various agents act differently based on the quality of the contagion they receive. We find that the spreading process is characterized by an initial phase where contagion spreads exponentially fast with time scales which strongly depend on the quality of the contagion being spread. This leads to a quasi-stationary state where extremely slow evolution eventually leads to a final equilibrium state. We note that such slow spreading occurring after the initial exponential growth has been observed in various contagion spreading phenomena before \cite{19,20}. However, the cause of such slow spreading has been associated with various types of correlations that may exist within standard models \cite{20}. In our model, however, both exponential and slow spreading exist with quality-dependent time-scales, and this general behavior is shown to hold on various network structures regardless of topology. Another important and often neglected feature of spreading is the activity of the agents. We therefore investigate such quantity and show that for a wide range of contagion quality, agent activities exhibits a power-law distribution, which again occurs independent of network structure. Such power-law behavior has recently been seen in real social networks \cite{19,20,21,22}.

Accordingly, we introduce a local quantity (called quality) $x_i$ for each node $i$. Here, we assume that the quality of each individual is directly proportional to the number of its neighbors $k_i$,
$x_i={k_i}/{k_{max}}$
where $k_{max}=max(k_i)$, making $0\leq x_i\leq 1$ for any given network. We also introduce a parameter $\alpha$ which characterizes the quality of the contagion being spread, where $0\leq\alpha\leq1$. Here, we consider a local probabilistic rule for acceptance (i.e. transmission) of contagion based on the perceived fitness of the incoming contagion, $\alpha$, which is defined as follows: for each node $i$ which receives the contagion $\alpha$, with probability $f(k_i,\alpha)=exp(-(x_i-\alpha)^2/2\sigma^2)$ accepts the contagion and with probability $1-f(k_i,\alpha)$ rejects it.  This is essentially our fitness criterion. Note that if the contagion is accepted, the individual keeps that contagion forever and passes it on to all its $k_i$ neighbors, otherwise the individual rejects the contagion. We note that other forms of stochastic as well as deterministic $\alpha$-dependent local rules may be considered \cite{18}. Here, we consider such Gaussian rule as we believe it represents a wider range of agent behavior. We propose to study such dynamics on both Erd\"{o}s-R\'{e}nyi (ER) \cite{1811} as well as scale-free (SF) \cite{1822} networks. Thus a randomly selected node is seeded with a contagion of quality $\alpha$ and the subsequent dynamics is monitored.

In order to take into account the heterogeneity induced by the presence of node with different connectivity, we consider the time evolution of the density of informed (infected) $I_k(t)$ and uninformed (uninfected) $U_k(t)$ nodes of connectivity $k$ at time $t$ \cite{3}. These variables are related by means of the normalization condition:
\begin{equation}\label{Eq.1}
    U_k(t)+I_k(t)=1
\end{equation}
The total density of the informed nodes in a network is expressed by an average over the various connectivity classes; i.e., $I(t)=\Sigma_kI_k(t)p(k)$, where $p(k)$ is the probability of a node having $k$ neighbors. These densities satisfy the following set of differential equations \cite{15,151}:

\begin{equation}\label{Eq.2}
\dfrac{\textrm{d}I_k(t)}{\textrm{d}t}=kU_k(t)f(k,\alpha)\Theta_k(t)
\end{equation}

 \begin{equation}\label{Eq.3}
\dfrac{\textrm{d}U_k(t)}{\textrm{d}t}=-kU_k(t)f(k,\alpha)\Theta_k(t)
\end{equation}
where the creation term is proportional to the node degree $k$, the probability $U_k(t)$ that a node with degree $k$ is not informed, the local probability of acceptance (transmission) $f(k,\alpha)$ and the density $\Theta_k(t)$ of the informed neighbors of a node of degree $k$ which received the contagion at time $t$. The last term is thus the average probability that any given neighbor of a node of degree $k$ is informed \emph{and} received that contagion at the exact time $t$. For uncorrelated networks, the probability that each link of an uninformed node is connected to an informed node of degree $k'$ is proportional to the fraction of links emanating from such nodes \cite{3}. By considering that at least one of the links of each informed node is connected to another informed node from which the contagion has been transmitted, one can show that $\Theta_k(t)=\Theta(t)$ \cite{4,41}, and in particular we obtain:
\begin{equation}\label{Eq.4}
\Theta(t)=\dfrac{\Sigma_{k'}(k'-1)p(k')I_{k'}(t)Q_{k'}(t)}{\left\langle k \right\rangle }
\end{equation}
\begin{equation}\label{Eq.5}
Q_k(t)=k\Theta(t-1)
\end{equation}
where $\left\langle k \right\rangle=\Sigma_{k'}k'p(k')$ is the proper normalization factor dictated by the total number of links and $Q_{k}(t)$ is the probability that a node of degree $k$ receives the contagion $\alpha$ at time $t$.

A reaction rate equation for $\Theta(t)$ can be obtained from Eqs. (3) and (5). In the initial spreading steps, one neglects terms of order $I_k^2$ and assuming, for early times, that the probability a node receives the contagion in its previous time step is one, leads to the following set of equations:
\begin{equation}\label{Eq.6}
\dfrac{\textrm{d}I_k(t)}{\textrm{d}t}=kf(k,\alpha)\Theta(t)
\end{equation}
\begin{equation}\label{Eq.7}
\dfrac{\textrm{d}\Theta(t)}{\textrm{d}t}=\dfrac{\Theta(t)}{\left\langle k \right\rangle} (\left\langle k^2f(k,\alpha) \right\rangle-\left\langle kf(k,\alpha) \right\rangle)
\end{equation}
These equations can be solved with the uniform initial condition $I_k(0)=I_0\ll1$ yielding for the total average density of informed nodes $I(t)=\Sigma_kp(k)I_k(t)$:
\begin{equation}\label{Eq.8}
I(t)=I_0[1+\tau \left\langle kf(k,\alpha) \right\rangle (\dfrac{\left\langle k \right\rangle-1}{\left\langle k \right\rangle})(e^{t/\tau}-1)]
\end{equation}
where
\begin{equation}\label{Eq.9}
\tau=\dfrac{\left\langle k \right\rangle}{\left\langle k^2f(k,\alpha) \right\rangle-\left\langle kf(k,\alpha) \right\rangle}
\end{equation}
is the growth time scale of an initial spreading in the network.

Eq.~(9) is a very interesting result since it highlights the importance of local dynamics in spreading through the appearance of $f(k_i,\alpha)$ assumed to be a general function which could be probabilistic or deterministic. Perhaps more importantly, for a given class of local dynamics (a given $f(k,\alpha)$) the spreading rate depends on the quality of the contagion, $\alpha$, being spread.

We also note that our results is a generalization of the standard (mean-field) SI model for contagion spreading, i.e., for a global transmission probability, $f(k,\alpha)=\lambda$, we obtain the standard result for heterogeneous mean-field models \cite{4}:
\begin{equation}\label{Eq.14}
\tau=\dfrac{\left\langle k \right\rangle}{\lambda(\left\langle k^2 \right\rangle - \left\langle k \right\rangle)}.
\end{equation}
On the other hand, for a homogeneous network, where nearly most nodes have a degree close to the mean degree $\left\langle k \right\rangle$, our results can be simplified by considering $\left\langle k^2 \right\rangle=\left\langle k \right\rangle^2+\left\langle k \right\rangle$ \cite{23}. Thus, the growth time scale in a homogeneous network can be obtained as:
\begin{equation}\label{Eq.15}
\tau\approx\dfrac{1}{\left\langle k \right\rangle f(\left\langle k \right\rangle ,\alpha)}
\end{equation}
which is to be compared with the usual $\tau=1/(\lambda \left\langle k \right\rangle)$ for homogeneous mean-field models \cite{4}.

\section{Numerical results}
We now present our simulation results for two types of random networks, SF with $p(k)\propto k^{-\gamma}$ with $\gamma=3$ and an ER network with a Poissonian distribution. The network realizations used for the numerical simulations were constructed using the method introduced in \cite{1833} in order to assure that no degree-degree correlations are present in any of the networks generated. We choose $\sigma=0.1$, and start by choosing a random site and infecting it with contagion $\alpha$. The reported results are typically averaged over 1000 seeding and a separate (ensemble) averaging of at least 10 realizations of a given random network. Fig.1 shows the time evolution of average density of infected/informed nodes $I(t)$ on both networks with the same average degree $\left\langle k\right\rangle =6$ and network size $N=20000$. Inset shows a log-linear plot of the early time behavior of $I(t)$ indicating an exponential growth with $\alpha$ dependent time scales. Fig.2 shows the growth time-scale $\tau$ obtained from exponential fitting of numerical results along with the theoretical prediction (Eq.9), versus $\alpha$. As can be seen, the numerical results recover the analytical calculation with great accuracy. We see that the fastest spreading (smallest $\tau$) coincides with largest spreading (largest $I$). However, as $\alpha$ deviates significantly from the average quality of the network, $\left\langle x\right\rangle $, $\tau$ increases significantly, and thus greatly slowing down the spreading process. We also note that low quality contagion is more efficiently spread on SF networks, which can be expected, as SF networks have low average quality $\left\langle x\right\rangle=\left\langle k\right\rangle/k_{max}$ because of the arbitrary large $k_{max}$.
\begin{figure}
\begin{center}$
\begin{array}{cc}
\includegraphics[width=8cm]{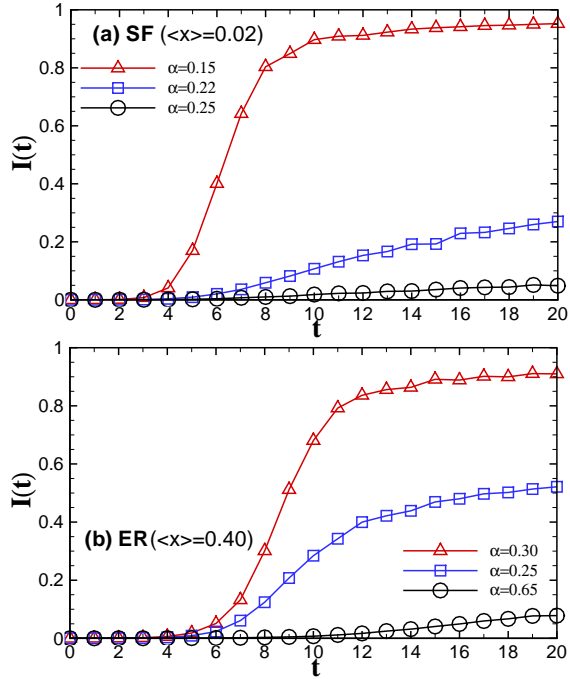}
\end{array}$
\end{center}
\caption{Averaged density of informed nodes versus time in (a) a SF network and (b) an ER network. The networks have the same size of $N=20000$ and average degree $\left\langle k\right\rangle=6$. The insets show the log-linear plot of the same data for the first few time steps. Note that the average quality $\left\langle x\right\rangle$ is much smaller on the SF network.}
\end{figure}

\begin{figure}
\begin{center}$
\begin{array}{c}
\includegraphics[width=8cm]{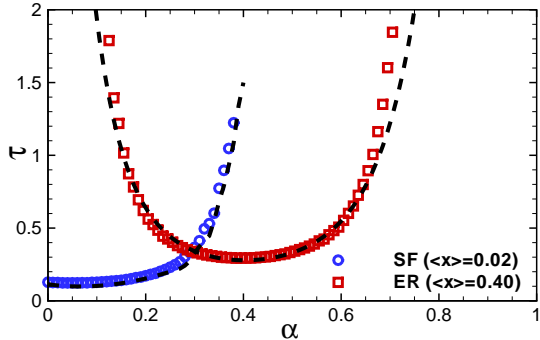}
\end{array}$
\end{center}
\caption{Measured time-scale $\tau$ vs $\alpha$ as obtained from exponential fitting in a SF network (circles) and an ER network (squares). Dashed lines indicate analytical calculation, Eq.(9). Parameters are the same as Fig.1}
\end{figure}

Based on the results shown in Fig.1, one might expect that initially fast spreading leads to a stationary state where a certain $\alpha$-dependent percentage of the nodes are informed. However, this is not the case. The spreading actually never stops in our model, and due to random nature of our local rules, even the nodes whose quality do not match well with the incoming contagion can eventually accept and thus transmit the contagion due to large number of exposures. This indicates that eventually all nodes are infected if one waits long enough. However, simulations show that the time required to reach such stationary state ($I(\infty)=1.0$) is very long and increases rapidly with system size $N$. What one typically observes after the fast exponential spreading is a quasi-stationary state where $\textrm{d}I/\textrm{d}t\approx0$ on short time scales, along with a very slow increase to the eventual stationary state. Of interest, is therefore, the profile of such quasi-stationary states as a function of $\alpha$. Fig.3 shows such results for various times. We observe that the most efficient spreading occurs when the quality of the contagion matches the average quality of the network, i.e. $\alpha=\left\langle x\right\rangle$. Significant spreading also occurs for a width of $2\sigma$ about such maximum. For a SF network (Fig.3(a)), average quality is nearly zero and the cutoff (at $\alpha=\left\langle x\right\rangle+2\sigma$) is relatively sharp. For ER network (Fig.3(b)), the average quality can differ, but spreading falls off gradually with the width $2\sigma$ about such an average. As different $\alpha$'s have different time scales, $\tau(\alpha)$, we also plot the profiles of quasi-stationary states after $100\tau(\alpha)$. This is shown in Fig.4 where the same type of behavior is seen as in Fig.3.

\begin{figure}
\begin{center}$
\begin{array}{ccc}
\includegraphics[width=8cm]{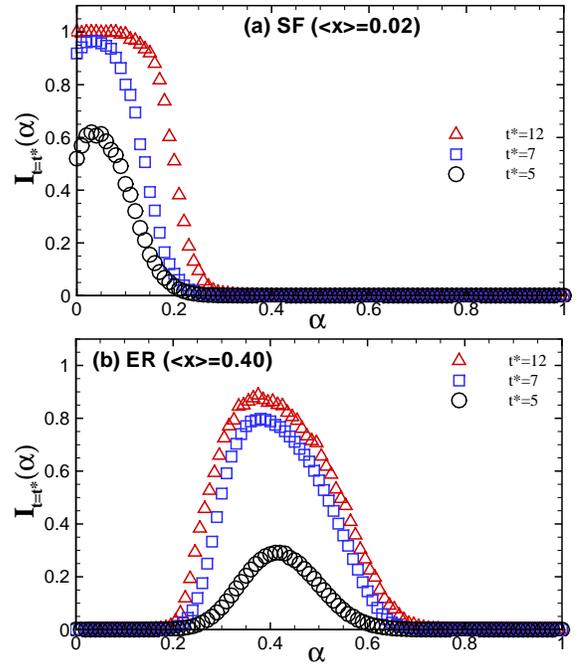}
\end{array}$
\end{center}
\caption{The density of informed nodes as a function of $\alpha$ for three different times on a (a) SF network and (b) ER network. Parameters are the same as Fig.~1.}
\end{figure}

\begin{figure}
\begin{center}$
\begin{array}{c}
\includegraphics[width=8cm]{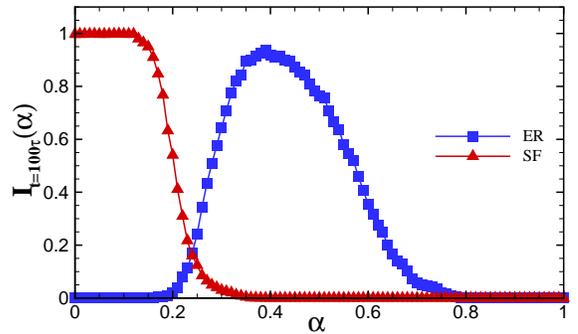}
\end{array}$
\end{center}
\caption{The density of informed nodes as a function of $\alpha$ at $t=100\tau(\alpha)$ on a SF network and ER network. Parameters are the same as Fig.~1.}
\end{figure}

We next ask how the eventual stationary state is reached. We find that the quasi-stationary state evolves with slow dynamics $I_{q.s.}\thickapprox t^{b(t)}$ with time-dependent exponents $b(t)=\sum_{j=2}^{\infty}c_j/(\log t)^j$ where $c_j$'s are $\alpha$ dependent. Fig.~5 shows the simulation results for various $\alpha$'s, along with a fit of the above function including the first three terms. Including higher order terms will make an increasingly better fit. Such multi-fractal slow dynamics to the eventual state is interesting in its own rights. The transition from fast initial exponential growth to the slow multifractal quasi-stationarity can be seen in Fig.1.

\begin{figure}
\begin{center}$
\begin{array}{ccc}
\includegraphics[width=8cm]{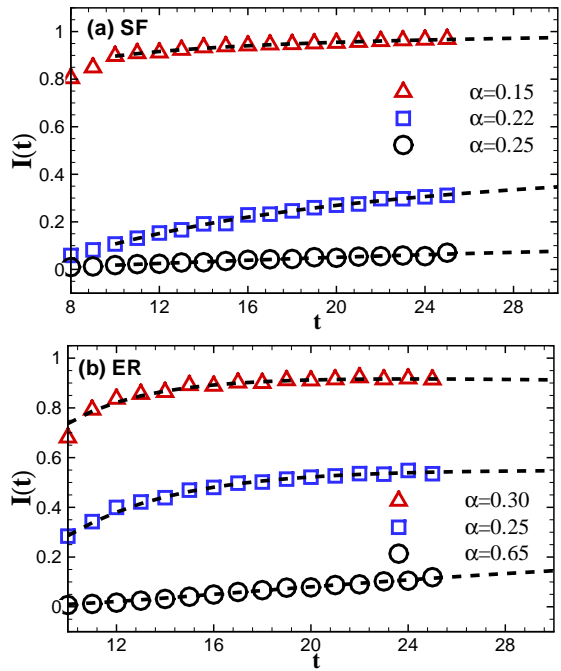}
\end{array}$
\end{center}
\caption{The long time behavior of the evolution of informed nodes on a (a) SF network of $\gamma=3$ and (b) ER network, both with the same mean degree $\left\langle k\right\rangle=15$. The dashed lines show the function $t^{b(t)}$ with $b(t)=\sum_{j=2}^{\infty}c_j/(\log t)^j$ where $c_j$ is $\alpha$ dependent fitting parameters and the fitting shows only the first three terms with $R^2\approx0.98$ while including higher order terms makes $R^2\rightarrow1$.}
\end{figure}

As noted earlier, spreading slows down greatly for values of $\alpha$ which significantly deviate from the average quality of the network $\left\langle x\right\rangle$, because of exponential growth of $\tau$, see Fig.2 and Eq.(11). This leads to a low value for $I_{q.s.}$ which can be interpreted as low prevalence. However, one can see that even if $\alpha$ does not significantly deviate from $\left\langle x\right\rangle$, a finite, arbitrary $\tau$ would lead, in a first few initial steps, to a finite number of infected/informed agents. It is therefore seen that low prevalence (i.e. small $I_{q.s.}$) could occur for large networks ($N\rightarrow\infty$) for generic values of $\alpha$. To show this we plot the value of $I_{q.s.}(t=20\tau)$ for $\alpha=\left\langle x\right\rangle+0.25$ ($\sigma=0.1$) for various system sizes. The results are shown in Fig.6. Interestingly, for both network types one observes a scaling $I_{q.s.}\sim N^{-\beta}$ indicating low prevalence for large system sizes. We note that the exponent $\beta$ is larger for SF networks than for ER networks showing a better tendency for low prevalence in SF networks. One might also wonder about the velocity $\textrm{d}I_{q.s.}(t)/\textrm{d}t$ as a function of $N$. Such results are shown in Fig.7, where the velocity also decreases (to zero) on SF networks while it remains at a constant (albeit small) value for ER networks, again showing a better tendency for low prevalence on SF networks.

\begin{figure}
\begin{center}$
\begin{array}{c}
\includegraphics[width=8cm]{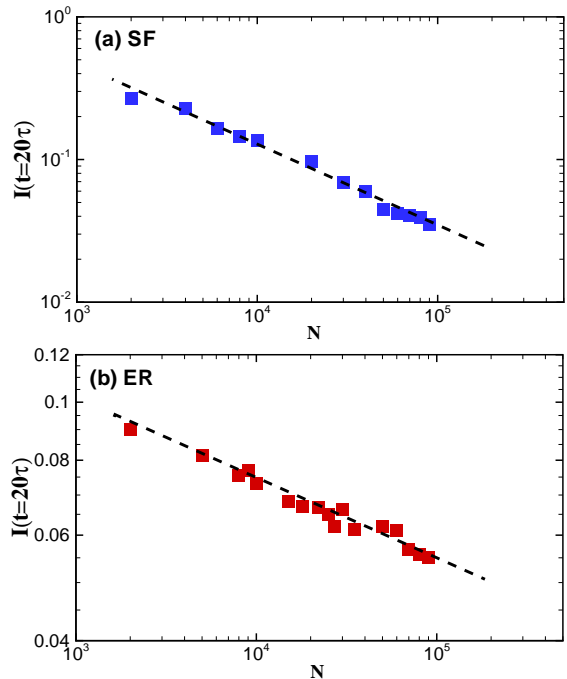}
\end{array}$
\end{center}
\caption{The size ($N$) dependence of the long time ($t\simeq20\tau$) value of the quasi-stationary state $I_{q.s.}=I(t\approx20\tau)$ for (a) a SF network and (b) an ER network for contagion $\alpha=\left\langle x\right\rangle+0.25$. The dashed line corresponds to $I_{q.s.}\sim N^{-\beta}$ with $\beta=0.57$ for SF (a) and $\beta=0.13$ for ER (b). The network parameters are as the same as in the Fig.1.}
\end{figure}

\begin{figure}
\begin{center}$
\begin{array}{c}
\includegraphics[width=8cm]{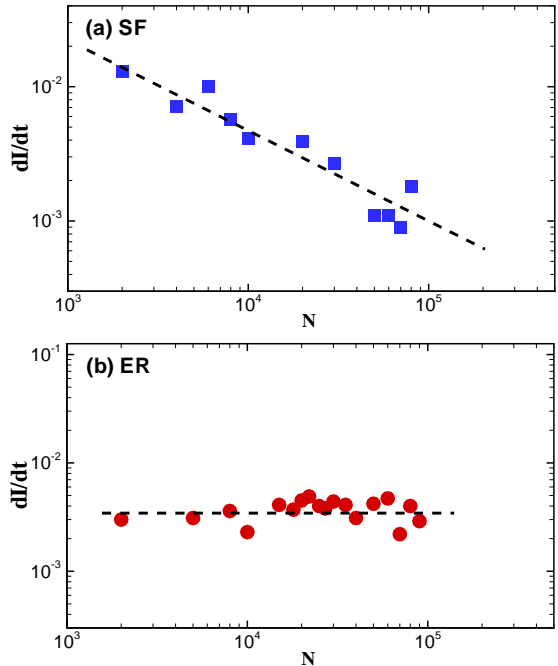}
\end{array}$
\end{center}
\caption{The velocity of the quasi-stationary evolution ($\textrm{d}I(t)/\textrm{d}t$) at long time ($t\simeq20\tau$) versus system size $N$ for (a) a SF network and (b) an ER network. The plot parameters are the same as in the Fig.1.}
\end{figure}

As noted above, nodes remain active in this model and can eventually activate previously inactive neighbors due to repeated transmissions. This is due to the probabilistic nature of local rules considered here. One can therefore wonder about the frequency of node's activity similar to how frequently users \textquotedblleft retweet\textquotedblright or \textquotedblleft comment\textquotedblright on a news in a social network [19,22]. We therefore define a node's activity as the number of times it has transmitted a given contagion, and plot the probability of such activity for large networks over long times. Fig.8 shows such results. Remarkably, for both types of networks and a wide range of contagion quality $\alpha$ (as long as the quality is not close to network average), a robust power-law behavior is observed, $P(a)\sim a^{-\delta}$ with $\delta=1.0$. Note that if $\alpha$ approaches $\left\langle x\right\rangle$ then a deviation from pure power-law is observed with the tail becoming more heavily weighted, where eventually for $\alpha=\left\langle x\right\rangle$ one observes a flat distribution with a strong and sharp peak in the tail. Power-law behavior in agents activities has been reported in various empirical studies  \cite{19,20,21,22}, with various exponents and various definitions of \textquotedblleft activity\textquotedblright. What seems to be important in our model is the generic conditions for such power-law behavior, i.e. a wide range of $\alpha$ and network topology, as well as the seemingly exact exponent $\delta=1.0$ over many decades.

\begin{figure}
\begin{center}$
\begin{array}{ccc}
\includegraphics[width=8cm]{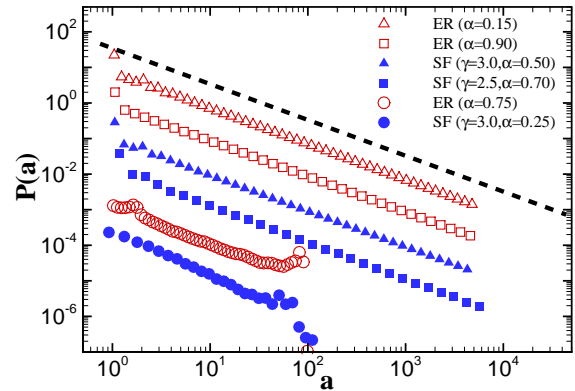}
\end{array}$
\end{center}
\caption{Distribution of node's activity for various contagion quality on both types of networks. The plots are shifted for clarity. Here, $N=20000$ and $\left\langle k\right\rangle=6$ with 50 seeds and 50 network realization for averaging. The dashed line is a line of slope $-1$ to help guide the eye.}
\end{figure}

\section{Concluding remarks}
In this work we have proposed a general model of spreading phenomena which takes into account the quality or strength of a given contagion. We also define a local probabilistic rule which models the fact that agents interact with contagions and transmit them based on their observed fitness. Our model reproduces some key features of spreading dynamics observed in empirical studies but absent in standard theoretical models. We find that the model exhibits quality-dependent fast initial spreading followed by a quasi-stationary state. The model also exhibits a robust power-law behavior of node's activities. Both these effects have been recently observed in empirical studies of social networks \cite{19,22}, leading us to believe that such a model, despite its simplicity, is an accurate description of spreading dynamics on complex networks. Another important feature of our model is the observed low prevalence as a result of large system sizes. For arbitrary values of contagion quality $\alpha$, and small $\sigma$, a finite number of agents are fit to receive the contagion and such fraction becomes ever smaller with increasing system sizes, more so on heterogeneous networks. For large enough system sizes, only contagions which fit the average network quality $\alpha\approx\left\langle x\right\rangle$ have any significant prevalence. We note that most naturally arising networks are heterogeneous in nature where $\left\langle x\right\rangle\rightarrow0$ in such limits. Therefore, this result may explain the curious popularity of tabloids, or the general prevalence of superstitious beliefs.

\begin{acknowledgments}
Support of Shiraz University Research Council is kindly acknowledged.
\end{acknowledgments}

\bibliographystyle{apsrev}

\end{document}